\documentclass[a4paper,fleqn,usenatbib]{mnras}

\usepackage[dvipdfmx]{graphicx}

\usepackage{amsmath}	% Advanced maths commands
\usepackage{amssymb}	% Extra maths symbols

\title[Photometric properties of the SDSS SNe Ia]{Photometric properties of intermediate redshift Type Ia Supernovae observed by SDSS-II Supernova Survey}
\author[N. Takanashi et al.]{N. Takanashi$^{1}$\thanks{E-mail:
naohiro.takanashi@emp.u-tokyo.ac.jp}, M. Doi$^{2,3,4}$, N. Yasuda$^{3}$, H. Kuncarayakti$^{5,6}$, K. Konishi$^{7}$, 
\newauthor D. P. Schneider$^{8,9}$, D. Cinabro$^{10}$, J. Marriner$^{11}$
\\
$^{1}$Executive Management Program, The University of Tokyo,  7-3-1 Hongo, Bunkyo-ku, Tokyo 113-8654, Japan.\\
$^{2}$Institute of Astronomy, Graduate School of Science, The University of Tokyo, 2-21-1 Osawa, Mitaka, Tokyo 181-0015, Japan.\\
$^{3}$Kavli Institute for the Physics and Mathematics of the Universe, The University of Tokyo, \\\phantom{0}5-1-5 Kashiwanoha, Kashiwa, Chiba 277-8568, Japan.\\
$^{4}$Research Center for the Early Universe, Graduate School of Science, The University of Tokyo, Bunkyo-ku, Tokyo 113-0033, Japan.\\
$^{5}$Millennium Institute of Astrophysics, Santiago, Chile.\\
$^{6}$Departamento de Astronom\'ia, Universidad de Chile, Casilla 36-D, Santiago, Chile.\\
$^{7}$Nikon Corporation, 471, Nagaodai-cho, Sakae-ku, Yokohama, Kanagawa 244-8533, Japan.\\
$^{8}$Department of Astronomy and Astrophysics, The Pennsylvania State University, University Park, PA 16802, USA.\\
$^{9}$Institute for Gravitation and the Cosmos, The Pennsylvania State University, University Park, PA 16802, USA.\\
$^{10}$Department of Physics and Astronomy, Wayne State University, Detroit, MI 48202, USA.\\
$^{11}$Center for Particle Astrophysics, Fermi National Accelerator Laboratory, Batavia, IL 60510, USA.\\
}

\date{Accepted 2016 October 20. Received 2016 October 19; in original form 2016 February 15}

\begin{document}

\pagerange{\pageref{firstpage}--\pageref{lastpage}} \pubyear{2014}

\maketitle

\label{firstpage}

\begin{abstract}

We have analyzed multi-band light curves of 328 intermediate redshift ($0.05 \le z < 0.24$) type Ia supernovae (SNe Ia) observed by the Sloan Digital Sky Survey-II Supernova Survey (SDSS-II SN Survey). The multi-band light curves were parameterized by using the Multi-band Stretch Method, which can simply parameterize light curve shapes and peak brightness without dust extinction models. We found that most of the SNe Ia which appeared in red host galaxies ($u - r > 2.5$) don't have a broad light curve width and the SNe Ia which appeared in blue host galaxies ($u - r < 2.0$) have a variety of light curve widths. The Kolmogorov-Smirnov test shows that the colour distribution of SNe Ia appeared in red / blue host galaxies is different (significance level of 99.9\%). We also investigate the extinction law of host galaxy dust. As a result, we find the value of $R_v$ derived from SNe Ia with medium light curve width is consistent with the standard Galactic value. On the other hand, the value of $R_v$ derived from SNe Ia that appeared in red host galaxies becomes significantly smaller. These results indicate that there may be two types of SNe Ia with different intrinsic colours, and they are obscured by host galaxy dust with two different properties. 

\end{abstract}


\section{INTRODUCTION}

Type Ia supernovae (SNe Ia) are known as excellent standard candles for tracing the expansion history of the Universe because of their homogeneity and high luminosity (\citealt{fil05}). Since the early 1990s, several SN Ia search programs which aimed to measure the Hubble constant have been performed. The CALAN/TOLOLO Supernova Search, started in 1990 \citep{ham93,phi93,ham95}, succeeded in producing a moderately distant ($0.01 < z < 0.10$) sample of SN Ia. The Supernova Cosmology Project (SCP) and the High-Z Team (HZT) are independent high redshift ($z \sim 0.5$) SN searches designed to measure the cosmological parameters of the Universe. After a careful analysis of their observed SNe Ia, both SCP and HZT reported the accelerating expansion of the Universe (\citealt{rie98,schm98,per99}). This result implies that there is an unknown component in the Universe, named ``dark energy''. 

As described above, SNe Ia show properties which are more homogeneous than those of any other astronomical objects which can be used as standard candles at cosmological distances. However, many studies have also revealed an intrinsic diversity of SN Ia properties. For example, there are several subgroups of SNe Ia which are classified by spectral features. SN1991T-type and SN1991bg-type are well studied subgroups of SNe Ia (\citealt{bra93,nug95}). The SN1991T-type SN Ia has a slightly broader light curve width and brighter peak magnitude than a normal SN Ia (\citealt{phi92}). The SN1991bg-type SN Ia has a narrower light curve width and fainter peak magnitude than a normal SN Ia (\citealt{fil92,lei93}). 

We can also find intrinsic diversity in spectra of SNe Ia without including any obviously SN1991T-like and SN1991bg-like SNe (e.g. \citealt{ben05}). Recent studies have reported that even the spectroscopically normal SN Ia (called ``Branch normal", \citealt{bra93}) may have subgroups. In order to describe the observed SN Ia rate, \cite{man05} proposed that the population of SNe Ia consists of two components, one named ``prompt'', whose rate is proportional to the star formation rate and is possibly related to a young progenitor, and the other named ``tardy'', whose rate is proportional to the total stellar mass of the host galaxy and is possibly related to an old progenitor (see also \citealt{sca05,man06}). Based on photometric and spectroscopic observations of SN2005hj, \cite{qui07} suggested that there may be two subgroups within Branch normal SNe Ia produced by two different progenitor channels. \cite{ell08} reported that rest frame ultraviolet spectra of SNe Ia have significant variation using a sample of 36 events at intermediate redshift ($z = 0.5$).

Recent studies with large SN Ia samples, from low redshift to high redshift, show a significant correlation between SNe Ia properties and their host galaxies. Using a local SN Ia sample, \cite{kel10} showed that SNe Ia occurring in physically larger, more massive hosts are $\sim10\%$ brighter after light curve correction. \cite{hic09b} demonstrated from CfA3 data that SNe Ia which appeared in Scd/Sd/Irr hosts and E/S0 hosts may arise from different populations (see also \citealt{nei09}). \cite{sul10} showed SNe Ia properties depend on the global characteristics of their host galaxies from SNLS and other data. \cite{lam10b} confirmed the effect of host galaxies on SNe Ia in the SDSS-II SN Survey. Also using data from the SDSS-II SN Survey, \citet{smith12} found that the rate of SN Ia per unit stellar mass is much higher in star-forming host galaxies compared to passive ones, and \citet{galbany12} suggested that SNe that exploded at large distances from their elliptical hosts tend to have narrower light curves. With SNFactory data, \citet{rig13} found possible two modes of SNe Ia. The first mode is present in all environments, and the second mode is exclusive to locally passive environments, intrinsically brighter,and occurs predominately in high-mass hosts. These relations between environment and propeties of SN Ia are used to the cosmological studies to reduce uncertainties in distance moduli (e.g. \citealt{rig15,jon15}). On the other hand, using by the SDSS-II SN Survey sample, \cite{wol16} reported there is no evidence of a significant correlation between Hubble residuals and multiple host galaxy properties such as host stellar mass and star formation rate. 

In this study, we investigate photometric properties of SNe Ia at intermediate redshift ($z \sim 0.2$) based on the previous study of photometric properties of 108 nearby ($z \sim 0.1$) SNe Ia in \cite{tak08} (hereafter, TAK08). We use data obtained by the SDSS-II SN Survey (\citealt{yor00,fri08}), which is well calibrated, homogeneous and the first large SN Ia sample at an intermediate redshift. The SDSS-II SN Survey is a SN survey with the wide-field SDSS 2.5-m telescope (\citealt{gun06}) and wide-field imaging camera (\citealt{gun98}) at Apache Point Observatory (APO) operated by an international collaboration. The SDSS-II SN Survey complements existing ground-based low redshift (LOTOSS, SNfactory, CSP etc.) and high redshift (ESSENCE, SNLS) SN search programs. Through repeated scans of the SDSS southern equatorial stripe (82N and 82S, about 2.5 deg wide by ~120 deg long) every other night over the course of three 3-month campaigns (Sept-Nov. 2005-7), the SDSS-II SN Survey obtained well-measured, densely sampled {\it u-, g-, r-, i-} and {\it z}-band light curves (\citealt{hol08, sak14}). 

In this paper, we present photometric properties of the SDSS SNe Ia parameterized by the Multi-band Stretch Method (TAK08). In particular, we focus on the stretch distribution of the SDSS SNe Ia and the relationship between a stretch factor and photometric properties. We first provide information about the data used in this work (\S2), then describe how we parameterize the multi-band light curves (\S3). The observational selection biases expected in the SDSS SNe Ia are discussed (\S4), and we show photometric properties of the SDSS SNe Ia, comparing them with those of nearby SN samples from the literature (\S5.1). We also examine properties of host galaxy dust (\S5.2).

The SNe Ia which were reported to the International Astronomical Union (IAU) have a SN ID like SN2005eg. However, there are many SNe Ia (most without spectroscopic typing configuration) that were not reported to the IAU in the SDSS sample used in this study. We use a SDSS SN ID like SDSS-SN12345 for the name of those SNe without IAU designations in this paper.

\section{PHOTOMETRIC DATA}

\begin{table*}
\begin{center}
\begin{minipage}{280pt}
\caption{Type and redshift of the SDSS SNe Ia.}
\label{SMP_DATA}
\end{minipage}
\end{center}
\begin{center}
\begin{tabular}{cccl} \hline\hline
Type & Number & Average Redshift & Comment \\ \hline
all & 328 & 0.17 & \\
SC & 239 & 0.16 & \\
SP & \phantom{0}18 & 0.17 &  \\
PP & \phantom{0}71 & 0.18 & Selected by \cite{sak08}\\
\hline
\end{tabular}\\
\vspace*{1.2pt}
{\footnotesize
\begin{minipage}{280pt}
SC is a spectroscopically confirmed SN Ia\\
SP is a spectroscopically probable SN Ia\\
PP is a photometrically probable SN Ia with spectroscopic redshift of the host galaxy \\
\end{minipage}}
\end{center}
\end{table*} %

In this study, we use {\it u-, g-, r-, i-} and {\it z}-band \citep{fuk96,doi10} light curves of 328 SNe Ia in the range $0.05 \le z < 0.24$\footnote{In order to reduce uncertainty of cross-filter K-correction, we use only SNe Ia at this redshift range where we do not to have to change a combination of filters.} obtained by the 2005-2007 SDSS-II SN Survey. As we show in Table \ref{SMP_DATA}, the SN Ia sample (hereafter SDSS sample) consists of three groups, which are (1) spectroscopically confirmed SNe Ia, (2) spectroscopically probable SNe Ia and (3) photometrically probable SNe Ia with host galaxy spectroscopic redshifts. The photometrically probable SNe Ia are identified as SNe Ia based on luminosity, colour, and light curve shapes (see details in \citealt{sak08}). We do not include known peculiar SNe Ia classified by \cite{zhe08} in the sample (SN2005js, SN2005gj, SDSS-SN7017, and SN2005hk are excluded). We do not use the SNe Ia whose photometry is not based upon the SDSS calibration stars in the same frame in order to avoid photometric systematic uncertainty. Also we do not use the SNe Ia which were observed after 5 days from the peak brightness in order to reduce uncertainties in estimating light curves width. Since we assigned higher priority of spectroscopic observations to not heavily extincted SNe Ia candidate (\citealt{sak08,sak14}), spectroscopically observed SNe Ia are biased to blue SNe Ia. To reduce the selection effect, we include photometrically probable SNe Ia in this study. Note that these photometrically probable SNe Ia are not spectroscopically confirmed as SNe Ia.

\begin{figure}
\includegraphics[scale=0.45]{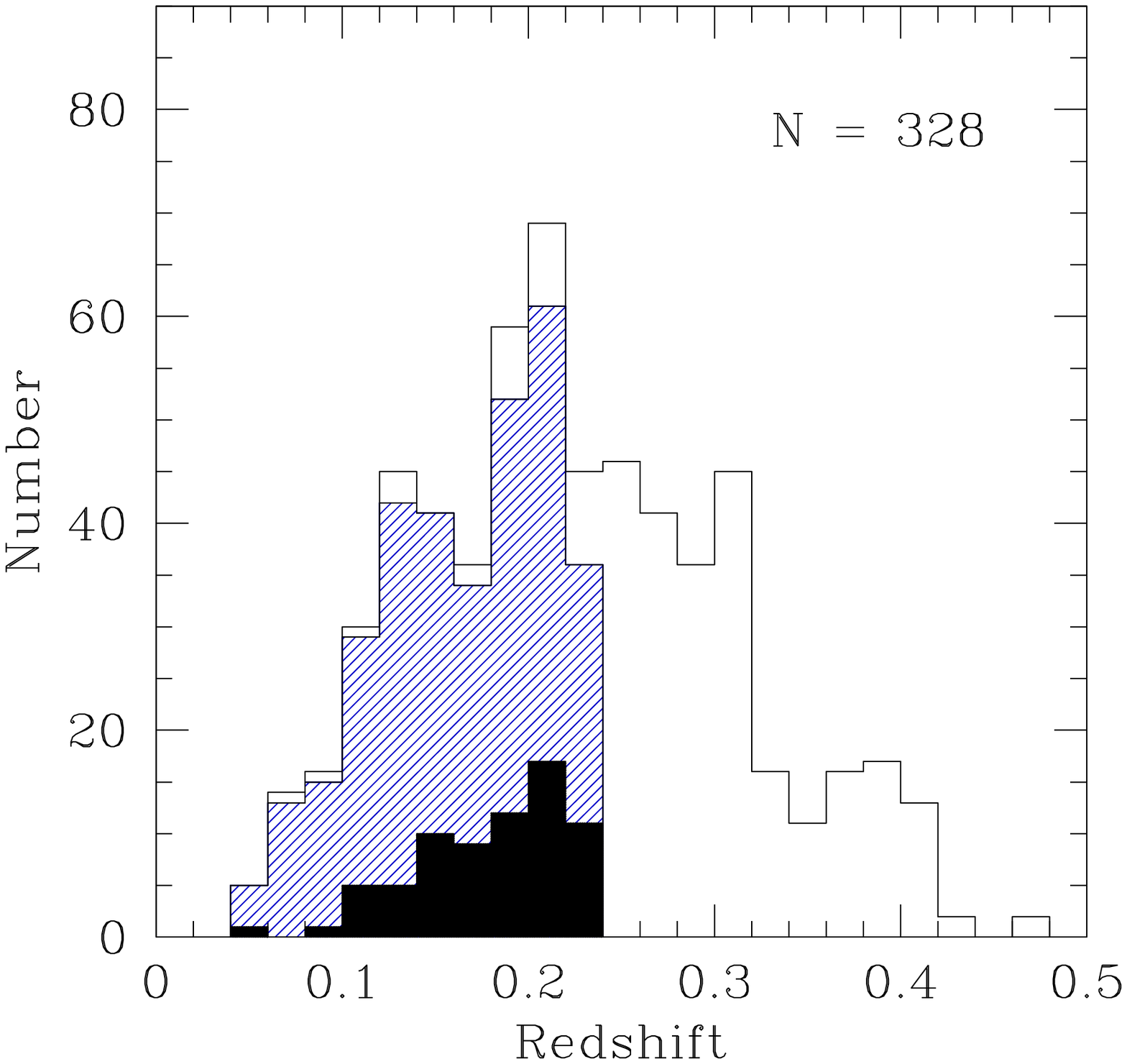}
\caption{Redshift distribution of the all SNe taken by the SDSS-II Supernova Survey. The shaded histogram shows the 328 SNe used in this work called ``the SDSS sample'' in this paper. Since we select the SN sample according to the criteria described in \S2, the number of the SDSS sample is not up to the all SNe sample even though at lower redshift. The filled histogram shows photometrically probable SNe Ia (see Table \ref{SMP_DATA}).  \label{z_HIST}}
\end{figure}

Figure \ref{z_HIST} shows the redshift distribution of the SDSS sample used in this study. In order to reduce uncertainty of cross-filter K-correction, we use only SNe Ia at this redshift range where we need not to change a combination of filters. The redshifts of all SNe in the sample are measured by spectroscopic observations of their host galaxies or SN spectra. The average redshift is $z \sim 0.17$, and about 60\% of samples lies in the range $0.15 < z < 0.24$. Because of insufficient telescope time for spectroscopic observations of faint objects, we did not obtain spectra for many SNe, especially those at higher redshift. We did acquire light curves of many of those missed ``SNe Ia", but we use only light curves which are classified as having a high probability of being a SN Ia (see \citealt{sak11}).

The photometry of the SNe Ia was mainly obtained by the SDSS camera on the 2.5-m telescope at APO. The flux of each SN Ia has been measured by \cite{hol08} with the technique named ``Scene Modeling Photometry (SMP)''. SMP does not use a template-subtracted image for photometry. The technique fits all of the individual reduced frames with a model of the galaxy background and the SN, and measures the flux from the model. The model of each image is generated from the sum of a set of stars, a galaxy, a SN, and background. As a result, SMP can avoid degrading the PSF and any spatial resampling that leads to correlated noise between pixels. The basic concept is similar to the technique of SNLS (\citealt{ast06}).

In order to discuss properties of the SDSS sample, we use the nearby SN Ia sample (hereafter Nearby sample) from TAK08. The Nearby sample includes 108 SNe Ia in the range $z < 0.11$ with {\it U-, B-, V-, R-} and {\it I-}band light curves. The light curves were parameterized by the Multi-band Stretch Method and the parameters are listed in Table 2 of TAK08.

\section{ANALYSIS}

We applied the ``Multi-band Stretch Method''  (TAK08) to the light curves in five passbands. This method is a revised Stretch method (cf. \citealt{per97,gol01}) extended to five bands. There are several methods for parameterizing light curve shapes of SN Ia. For example, ``MLCS'' \citep{rie96} and ``SALT2'' \citep{guy07} are widely used algorithms. These methods fit parameters of light curve width and dust extinction to observed light curves at once. The reason that we apply a stretch factor to {\it U-, B-, V-, R-} and {\it I-}band independently instead of a common stretch factor like SALT2 is that we want to investigate the diversity of light curve shapes independently. We can simply parameterize light curve shapes and peak brightness without any additional assumptions such as dust extinction models.

We fit up to 11 parameters to each event: {\it U-, B-, V-, R-} and {\it I-}band stretch factors, {\it U-, B-, V-, R-} and {\it I-}band peak magnitudes, and a time ${\it t_{Bmax}}$ for maximum light in the {\it B-}band. Only ${\it t_{Bmax}}$ is common to all passband, and the remaining 10 parameters are treated equally and independently. The method does not include any dust / colour corrections and has the advantage that we can simply parameterize the light curve shape and luminosity without any assumptions. We used the {\it U-, B-, V-, R-} and {\it I-}band light curve templates which are adjusted to coincide with a SN Ia with $s_{(B)} = 1.0$ SED template (\citealt{nug02}) for the light curve fitting. We derived absolute magnitudes for the SDSS sample using redshifts under the assumption of the flat $\Lambda$CDM model with standard cosmological parameters ($H_{0} = 68.5$ km/s/Mpc$, \Omega_{M} = 0.303, \Omega_{\Lambda} = 0.697$ from \citealt{bet14}). We corrected extinction by dust in the Milky Way according to \cite{sch98}. We also corrected cosmological time dilation. The differences from the analysis in TAK08 are that (1) we need to transform the photometric data from the native SDSS system to the Vega system, and (2) we must apply cross filter K-corrections to the data. We describe these procedures below.

\subsection{Magnitude System}

The SMP data are provided in the natural system of the SDSS 2.5-metre telescope (\citealt{doi10}). We need to transform this ugriz data to the Vega system to compare SDSS photometry with {\it U-, B-, V-, R-} and {\it I-}band light curves of the Nearby sample. We transformed the data in the following way.

\begin{table}
\begin{center}
\begin{minipage}{160pt}
\caption{Synthetic AB magnitude of Vega in the SDSS filters. \vspace{4pt}}
\label{VEGA_AB_MAG}
\end{minipage} 
\begin{tabular}{ccccc} \hline\hline
{\it u} & {\it g} & {\it r} & {\it i} & {\it z} \\ \hline
 0.951 & -0.080 & 0.169 & 0.389 & 0.556 \\ \hline
\end{tabular} \\
{\footnotesize
\begin{minipage}{160pt}
\vspace*{1.2pt}
Calculated from the STSDAS v3.3 synphot Vega spectrum assuming V=+0.03.\vspace{4pt}
\end{minipage}}
\end{center}
\end{table} %

First, the native SDSS system was transformed into the AB system (\citealt{sak14}). 

\begin{eqnarray}
u(AB) & = u(SDSS) - 0.0679 \nonumber\\
g(AB) & = g(SDSS) + 0.0203 \nonumber\\
r(AB) & = r(SDSS) + 0.0049 \nonumber\\
i(AB) & = i(SDSS) + 0.0178 \nonumber\\
z(AB) & = z(SDSS) + 0.0102 \nonumber
\end{eqnarray}

Next, the AB system was transformed into the Vega system. To adjust zeropoints of the AB system and the Vega system, the {\it u-, g-, r-, i-} and {\it z}-band magnitude of Vega in the AB system (Table \ref{VEGA_AB_MAG}) were added as follows.

\begin{eqnarray}
u(Vega) & = u(AB) - u(AB)_{Vega}\nonumber\\
g(Vega) & = g(AB) - g(AB)_{Vega}\nonumber\\
r(Vega) & = r(AB) - r(AB)_{Vega}\nonumber\\
i(Vega) & = i(AB) - i(AB)_{Vega}\nonumber\\
z(Vega) & = z(AB) - z(AB)_{Vega}\nonumber
\end{eqnarray}

At the end of this procedure, we have transformed the {\it u-, g-, r-, i-} and {\it z}-band magnitudes of the SDSS SNe into the Vega system.

\subsection{Cross-Filter K-correction}

\begin{figure}
\begin{center}
\includegraphics[scale=0.45]{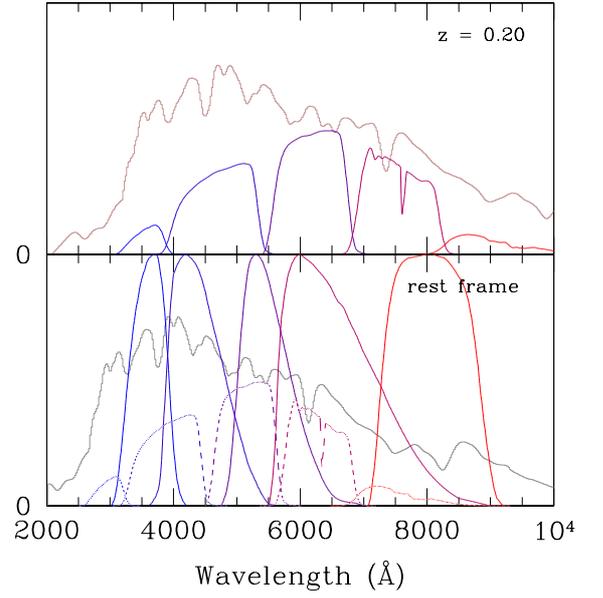}
\end{center}
\caption{In the top panel, we show {\it u-, g-, r-, i-} and {\it z}-band response curves (including atmosphere) and a SN Ia spectrum at maximum which is shifted to $z = 0.2$. In the bottom panel, we show {\it U-, B-, V-, R-} and {\it I-}band responses in solid lines, shifted {\it u-, g-, r-, i-} and {\it z}-band response at $z = 0.2$ in dashed lines, and a rest frame SN Ia spectrum at maximum. \label{FILTER}}
\end{figure}

The next step is to apply K-corrections to transform observed {\it u-, g-, r-, i-} and {\it z}-band magnitudes to {\it U-, B-, V-, R-} and {\it I-}band magnitudes (see Figure \ref{FILTER}). We have to choose an appropriate filter combination for each cross-filter K-correction as a function of redshift. In this study, we transformed ({\it u, g, r, i, z}) $\rightarrow$ ({\it U, B, V, R, I}) in the range $0.05 \le z < 0.24$.

The cross-filter K-correction was applied according to \cite{kim96}. We use the spectrum of \cite{hsi07} for the K-correction. Ideally, we should apply a spectrum template based on the proper type of each SN Ia (e.g. SN1991T-like, SN1991bg-like) for K-corrections, but we applied Hsiao's template all SNe Ia. The template is warped to fit the observed ugriz photometry. We use the standard galactic extinction curve (\citealt{car89}) for warping.

\begin{table*}
\begin{center}
\begin{minipage}{240pt}
\caption{Relations between {\it B-}band stretch factor and stretch factors in {\it U-} and {\it V-}band of the Nearby SNe Ia from TAK08}
\label{Bsf_sf_tab}
\end{minipage} \\
\small
\begin{tabular}{ccc} \hline\hline
Relation & r.m.s. & Number \\ \hline
$s_{(U)} = (0.95 \pm 0.02) \times s_{(B)} + (0.16 \pm 0.02)$ & 0.09 &  75 \\ 
$s_{(V)} = (0.88 \pm 0.01) \times s_{(B)} + (0.16 \pm 0.01)$ & 0.08 &  92 \\ \hline
\end{tabular}
\end{center}
\end{table*}

\begin{figure}
\begin{center}
\includegraphics[scale=0.45]{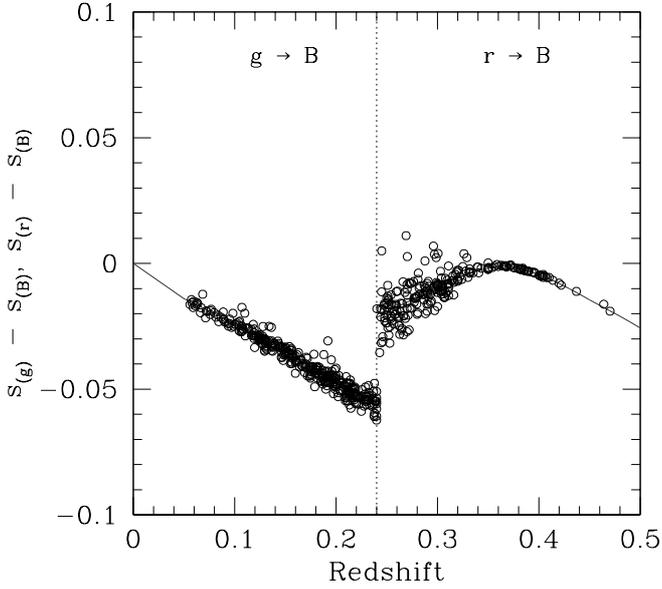}
\end{center}
\caption{Stretch corrections of each SN Ia versus redshift. We need to correct the {\it g-}band and {\it r-}band stretch factor to the rest frame {\it B-}band stretch factor ({\it g-}band in the range $z \le 0.24$ and from {\it r-}band in the range $0.24 < z < 0.50$) since stretch factors of a SN Ia in different filter bands are not same. The solid line shows the size of stretch correction of SN Ia ($s_{(B)} = 1.0$). See text for description. \label{Z_BSFCRR}}
\end{figure}

Since we have calibrated the flux with only the cross-filter K-correction, we should also apply the stretch K-correction. As we have noted above, we chose the appropriate filter combination for the cross-filter K-correction as a function of redshift, but the K-corrected bandpass of observed band is not the same as that of rest frame {\it B-}band. We should correct the difference of stretch factors due to the different bandpass. We use the relations between $s_{(B)}$ and the stretch factors of other bands derived from the Nearby SNe Ia (Table \ref{Bsf_sf_tab}) for the stretch K-correction. Based on those relations, we calculated the size of the stretch K-correction by interpolating stretch factors of different two bands. Figure \ref{Z_BSFCRR} shows the size of stretch corrections of the SDSS SNe Ia versus redshift. We show the typical size of the stretch correction of the SN Ia with $s_{(B)} = 1.0$ as a solid line. The correction becomes smallest at the redshift where the K-corrected bandpass of observed band agrees well with that of rest frame {\it B-}band. For the SDSS SNe Ia, the correction reaches a maximum of 5\% in the range $0.05 \le z < 0.24$.

\subsection{Uncertainty of the Analysis}

We now consider the uncertainties in our analysis. The uncertainties are divided into three components: uncertainty due to the light curve fitting (including observational photometric uncertainties), uncertainty due to the K-correction, and uncertainty due to the cosmological parameters adopted. 

We estimated the uncertainty due to the light curve fitting with Monte Carlo simulations. We created artificial light curves from the light curve template based on the observed dates and photometric uncertainties of each epoch. We measured the dispersion of each fitting parameter (luminosity, stretch factor, and time), and estimated the size of uncertainties. The typical uncertainties in estimating luminosity at $z = 0.2$ are $\sim 0.4$ mag in {\it U-}band, $\sim 0.06$ mag in {\it B-, V-} and {\it R-}band, and $\sim 0.2$ mag in {\it I-}band. The typical uncertainties in estimating stretch factors at $z = 0.2$ are $\sim 0.3$ in {\it U-}band, $\sim 0.05$ in {\it B-} and {\it V-}band, $\sim 0.1$ in {\it R-}band, and $\sim 0.3$ in {\it I-}band.

\begin{figure}
\begin{center}
\includegraphics[scale=0.45]{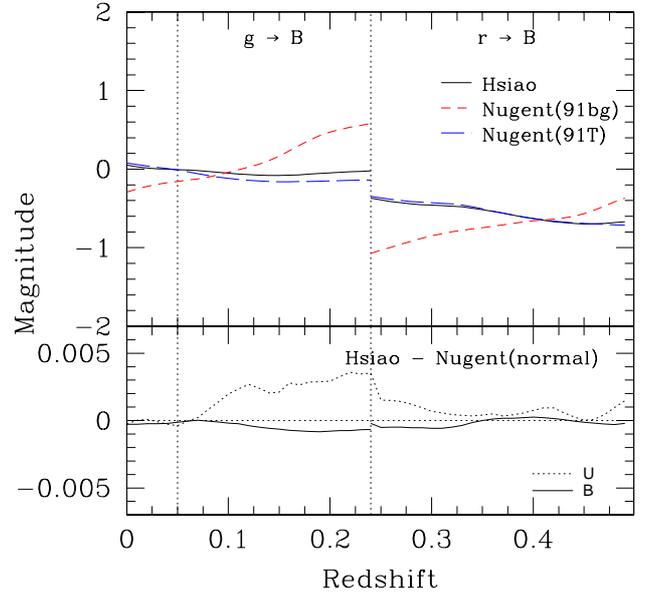}
\end{center}
\caption{In the top panel, we show the amount of K-correction (${\it g} \rightarrow {\it B}, {\it r} \rightarrow {\it B}$) at {\it B-}band maximum calculated from three different types of SN Ia template versus redshift. The black solid line is calculated from Hsiao's template (corresponding to Nugent's normal template), the short dashed red line is calculated from Nugent's SN1991bg-like template, and the long dashed red line is calculated from Nugent's SN1991T-like template. In the bottom panel, we show the difference between K-corrections which calculated from Hsiao's template and Nugent's Branch Normal template in {\it U-} and {\it B-}band versus redshift. Vertical dotted lines denote the border of band combinations for K-correction. \label{Z_KCRR}}
\end{figure}

It is difficult to estimate the uncertainty due to the K-correction. As described above, we applied Hsiao's template to all SNe Ia in our sample, some SNe in our sample might include SNe Ia which have spectra which differ from Hsiao's template. If one particular SN Ia has a spectrum which differs strongly from Hsiao's template such as that of SN1991T or SN1991bg, we might significantly overestimate or underestimate the size of K-correction. For example, we show the value of K-correction at {\it B-}band maximum versus redshift in the top panel of Figure \ref{Z_KCRR}. As is clear in the figure, the difference is significantly larger than the uncertainties due to light curve fitting. 

If the SN Ia has a spectrum similar to that of Hsiao's template, the typical uncertainties in K-correction is 0.01 to 0.02 mag in the range $z < 1.0$ from SNLS study (\citealt{fol08b}). We also compared the difference between Hsiao's template and Nugent's template ``Branch Normal" (in the bottom panel of Figure \ref{Z_KCRR}), and confirmed the difference is negligible (the size is only $< 0.005$ mag in {\it U-} and {\it B-}band).

The uncertainty due to the cosmological parameters also affects our estimation of distance modulus (absolute magnitude). If $H_0$ moves 1$\sigma$ ($\pm 1.27$ km s$^{-1}$Mpc$^{-1}$, \citealt{bet14}), the distance modulus changes by 0.04 mag. Adopting a slightly different value for $H_0$ only shifts the zeropoint of the absolute magnitude, and it does not affect the conclusions in this study. A change of 1$\sigma$ in $\Omega_{m}$ and $\Omega_{\Lambda}$ is equivalent to only 0.002 mag of distance modulus at $z = 0.2$. Note that the uncertainty is negligible for the comparison of samples with the same redshift, but is important for the comparison of samples with different redshifts.

Any uncertainty in the redshift of an event creates a corresponding uncertainty in its distance modulus. The redshifts of all SNe Ia used in this study were measured by a host galaxy spectrum or a SN spectrum, with a typical uncertainty in redshift of 0.001; this corresponds to an uncertainty of 0.01 mag at $z = 0.2$.

The main component of uncertainty is due to light curve fitting (see Table \ref{ERROR_TABLE}). In the following discussion, the uncertainties of each parameter include all of uncertainties.

\begin{table*}
\begin{center}
\begin{minipage}{350pt}
\caption{Typical uncertainty in estimating luminosity and stretch factors at $z = 0.2$. \vspace{4pt}}
\label{ERROR_TABLE}
\end{minipage}
\begin{tabular}{lccl} \hline\hline
 & $\sigma_{M_{B}}$ & s.f. & Comment \\ \hline
Light Curve Fitting & $0.06$ & $0.05$ & including photometric uncertainties of each epoch \\ 
K-correction & $0.02$ & - & Branch Normal SN Ia only, \cite{fol08a}\\ 
Cosmological Parameters & $0.002$ & - &  \\ 
Redshift Estimation & $0.01$ & - &  \\  \hline
Total & $0.06$ & $0.05$ &  \\ \hline 
\end{tabular}
\end{center}
\end{table*}

\section{OBSERVATIONAL BIAS OF THE SDSS SAMPLE}

In this section, we discuss the observational biases of the SDSS sample used in this study.

\subsection{Selection Bias of the SDSS Sample}

\begin{figure}
\begin{center}
\includegraphics[scale=0.45]{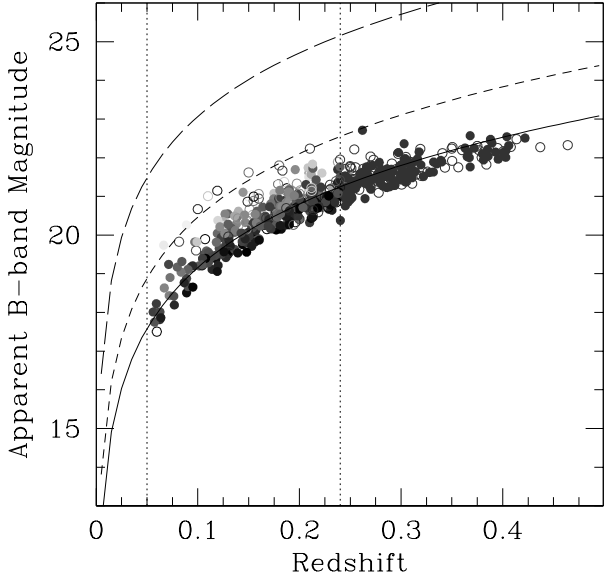}
\end{center}
\caption{Apparent peak {\it B-}band magnitude versus redshift of the SNe Ia observed by the SDSS-II Supernova Survey with uncertainties less than 0.2 magnitude. Filled symbols are spectroscopically confirmed SNe Ia and open symbols are spectroscopically or photometrically probable SNe Ia. The solid line denotes typical luminosity of SN Ia, -19.2 + distance modulus. The dashed lines denote the limits of host galaxy dust extinction parameter $A_{V}$ (we transformed $A_{V}$ to $A_{B}$ in this figure); we set the value to 1.0 mag for the 2005 survey (short dashed line) and 3.0 mag for the 2006-2007 survey (long dashed line). SNe plotted between dotted lines are the SDSS sample. The darker symbols are the bluer SNe Ia and the lighter symbols are the redder SNe Ia.
\label{Z_B}}
\end{figure}

The most important case of selection biases in the SDSS sample is due to the detection limit of the SDSS 2.5-metre photometry in which the candidates are identified (see Figure \ref{Z_B}). The signal-to-noise ratio (S/N) threshold for object detection is $\sim 3.5$, in typical conditions, corresponding to $g \sim 23.2, r \sim 22.8,$ and $i \sim 21.2$ mag in AB system (\citealt{dil08}). From Monte-Carlo simulations with artificial SNe, \cite{dil08} showed the survey efficiency is 100\% for candidates whose {\it g}-band maximum brightness is brighter than $\sim 21.2$ mag.

Another aspect is the bias due to the SN search strategy, especially spectroscopic target selection (see \citealt{sak08}). Since the survey typically found 10 SN candidates per night, they could not all be observed spectroscopically. We chose targets for spectroscopic observations based on their probability of being SNe Ia. The likelihood was calculated from {\it g-, r-} and {\it i}-band light curve fitting with four parameters, $z, A_{Vsdss}, T_{max},$ and a $template$;  here $A_{Vsdss}$ is the host galaxy extinction in {\it V-}band under the assumption that $R_{V} = 3.1$, $T_{max}$ is the time of rest-frame {\it B-}band maximum light, and the $template$ is taken from a set of seven light curves, of SNe type Ia (normal, SN1991T-like, and SN1991bg-like), Ib, Ic, II-P and II-L SNe. 

We searched for best fit parameters for each candidate, and ranked them by the probability of being SNe Ia. Highly extincted SNe (beyond $A_V$ = 1.0 for 2005 run and $A_V$ = 3.0 for 2006, 2007 run) are removed from the SDSS sample, but number of those missed SNe Ia may be very small (\citealt{sak08} reported only 2 SNe are missed in 2005 run, see also Figure \ref{Z_B}). Many SNe Ia with lower priorities were undoubtedly lost from spectroscopically confirmed SNe Ia subsamples, especially if the colour of SN Ia differs from the $template$ or the host galaxy extinction law differs from the galactic extinction law. In fact, the average $(M_B-M_V)_{max}$ of spectroscopically confirmed SNe Ia is 0.05, which is bluer than those of spectroscopically or photometrically probable SNe Ia (average $(M_B-M_V)_{max} = 0.09$). In order to reduce the selection bias due to spectroscopic target selection, the SDSS sample used in this paper includes not only spectroscopically confirmed SNe Ia  but also spectroscopically or photometrically probable SNe Ia. 

\subsection{Contamination of other type SNe}

As discussed in \S2, we added spectroscopically or photometrically probable SNe Ia to the SDSS sample in order to reduce the biases described above and to preserve the diversity of the sample since spectroscopically confirmed SNe Ia are strongly biased (\citealt{sak11}). However, the incorporation of photometrically probable SNe Ia may introduce contamination by other types of SNe. According to spectroscopic observations of SN Ia candidates which correspond to photometrically probable SNe Ia, the probability of misidentification is about 10\% (\citealt{dil08}). Since the SN candidates were selected based only on photometry at early phases, the probability of misidentification would be smaller than 10\% for the photometrically probable SNe Ia with late epoch observations. \cite{sak11} estimates purity of the photometrically probable SNe Ia is $\sim$91\% and contamination of other type SN is $\sim$6\%. So the number of those misidentified SN must be less than 7 in the SDSS sample (we have 71 photometrically probable SNe Ia at the redshift range). 

\begin{table*}
\begin{center}
\begin{minipage}{400pt}
\caption{The sample of the fitting parameters of the 328 intermediate redshift SDSS SNe Ia. Complete data are provided at online.} \label{data}
\end{minipage} \\
\begin{tabular}{cccccccc} \hline\hline
SDSS-SNID & SN name & redshift & $s_{(B)}$ & $M_{B}$ & $M_{V}$ & $M_{R}$ & $M_{I}$ \\ \hline
762 & 2005eg & 0.19 & 1.14(0.05) & -18.86(0.07) & -18.89(0.03) & -18.88(0.05) & -18.58(0.32)\\
1032 & 2005ez & 0.13 & 0.74(0.03) & -18.42(0.11) & -18.49(0.05) & -18.62(0.04) & -18.77(0.16)\\
1241 & 2005ff & 0.09 & 0.93(0.01) & -18.83(0.04) & -18.91(0.03) & -18.93(0.03) & -18.78(0.07)\\
1371 & 2005fh & 0.12 & 1.09(0.02) & -19.47(0.04) & -19.38(0.03) & -19.35(0.04) & -19.17(0.06)\\
1395 & - & 0.19 & 1.00(0.06) & -18.77(0.10) & -18.75(0.06) & -18.92(0.06) & -18.73(0.16)\\
1415 & - & 0.21 & 1.12(0.08) & -18.51(0.07) & -18.78(0.05) & -18.79(0.06) & -19.06(0.20)\\
1525 & - & 0.11 & 1.73(0.08) & -18.41(0.05) & -18.28(0.02) & -18.65(0.03) & -18.48(0.03)\\
1595 & - & 0.21 & 1.00(0.05) & -19.10(0.07) & -19.05(0.06) & -19.10(0.05) & -18.91(0.17)\\
1740 & - & 0.17 & 0.94(0.05) & -18.80(0.08) & -18.79(0.03) & -18.86(0.04) & -18.79(0.16)\\
1794 & 2005fj & 0.14 & 1.21(0.06) & -18.85(0.05) & -18.87(0.03) & -18.84(0.03) & -18.52(0.19\\ \hline
\end{tabular} \\
\vspace*{0.6em}
{\footnotesize
\begin{minipage}{400pt}
Peak magnitudes are shown in Vega system. Redshift of the SNe Ia is published in \cite{sak11}. \end{minipage}}
\end{center}
\end{table*}

\section{RESULTS}

\subsection{Photometric Properties of the SDSS sample}

As we described above, we use 328 SNe Ia in the range $0.05 \le z < 0.24$. At this redshift range, the bluer SNe Ia are almost complete, but the redder SNe Ia are not complete (Figure \ref{Z_B}). However, the selection bias due to the redshift cutoff is not important for the following discussion. 

Because of poor {\it u-} and {\it z}-band photometry relative to {\it g-, r-} and {\it i}-band, we do not use stretch factors and peak magnitudes of {\it u-} and {\it z}-band in this study ({\it u}-band corresponds to {\it U-}band and {\it z}-band corresponds to {\it I-}band in the range $z < 0.24$). Also the sample does not include SNe with large photometric uncertainties ($> 20\%$). All parameters of the 328 SNe Ia are given at the online table (partly given in Table \ref{data}).

\subsubsection{Stretch Factor Distribution}

\begin{figure}
\begin{center}
\includegraphics[scale=0.45]{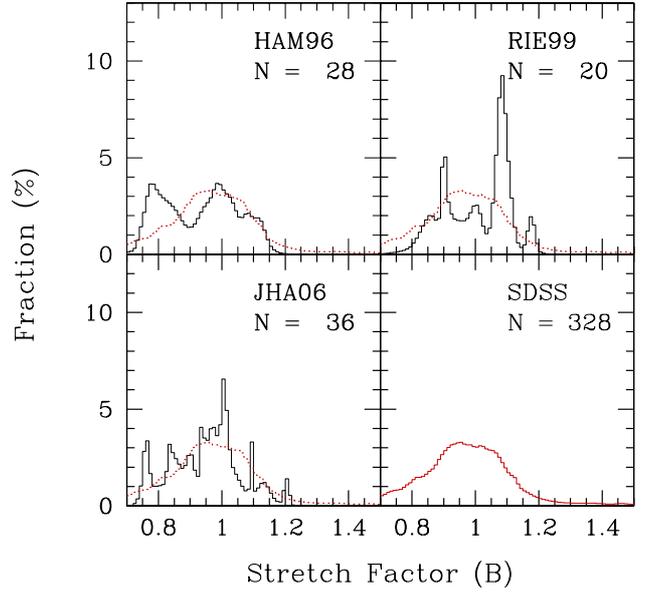}
\end{center}
\caption{Distribution of {\it B-}band stretch factors of the Nearby SNe Ia (TAK08) found by three different groups (HAM96, RIE99 and JHA06). Black line is the Nearby SN Ia sample found by each group, red line is the distribution of the SDSS SNe Ia in the range $0.05 \leq z < 0.24$ and red dotted line is the same distribution of the SDSS sample. The histograms are made from {\it B-}band stretch factors which are smoothed by estimated uncertainties. \label{BSF_HIST_GAUSS_COMP}}
\end{figure}

The stretch factor is one of the simplest parameters reflecting intrinsic properties of SNe Ia among the parameters derived from light curve fitting since stretch factors are not affected by the host galaxy dust. In the lower right panel of Figure \ref{BSF_HIST_GAUSS_COMP}, we show the distribution of {\it B-}band stretch factor of the SDSS sample. In this figure, we also show the stretch distributions of the Nearby SNe Ia used in TAK08, in which three major data sets were identified (\citealt{ham96,rie99b,jha06}, hereafter HAM96, RIE99, JHA06; 28 SNe Ia from HAM96, 20 SNe Ia from RIE99, 36 SNe Ia from JHA06). 

The SNe Ia shown in RIE99 and JHA06 were not discovered in a systematic SN search program. Their SNe Ia were selected from the IAU's Central Bureau for Astronomical Telegrams (CBAT). On the other hand, the SNe Ia shown in HAM96 were discovered by the CALAN/TOLOLO Supernova Search (\citealt{ham93}), which is a repeated scan SN survey of galaxy clusters which include many early type galaxies. This fact may result in an excess of events with a lower stretch value in the HAM96 sample since early type galaxies tend to host lower stretch SNe Ia (c.f. \citealt{how01,van05}). The stretch distributions of HAM96 shown in Figure \ref{BSF_HIST_GAUSS_COMP} seem to have more SNe Ia with small stretch values than those in RIE99 and JHA06; recall that SNe with small stretch values are expected to be fainter than average SNe Ia. This difference may be caused by observational selection biases since the samples made from the SNe Ia reported in CBAT are expected to include fewer less-luminous SNe Ia with narrower light curves.

In order to compare photometric properties of the SDSS SNe with different stretch factors, we classified the SDSS sample into three subsamples as follows; (1) ``Narrow" : $s_{(B)} \le 0.9$, (2) ``Medium" : $0.9 < s_{(B)} \le 1.1$, and (3) ``Broad" : $s_{(B)} > 1.1$. 

\subsubsection{Stretch-Magnitude Relation}

We compared the stretch-magnitude relations derived from the bluest SNe ($-0.14 \le (M_B-M_V)_{max} \le -0.10$) of the Nearby and SDSS samples (Table \ref{Bsf_mag_coe}), which are expected to be almost free from dust extinction. Slope of the {\it B-}band stretch-magnitude relation of the SDSS sample and the Nearby sample are consistent within the uncertainties, though that of the {\it V-}band stretch-magnitude relation of them are not consistent.

\begin{figure}
\begin{center}
\includegraphics[scale=0.45]{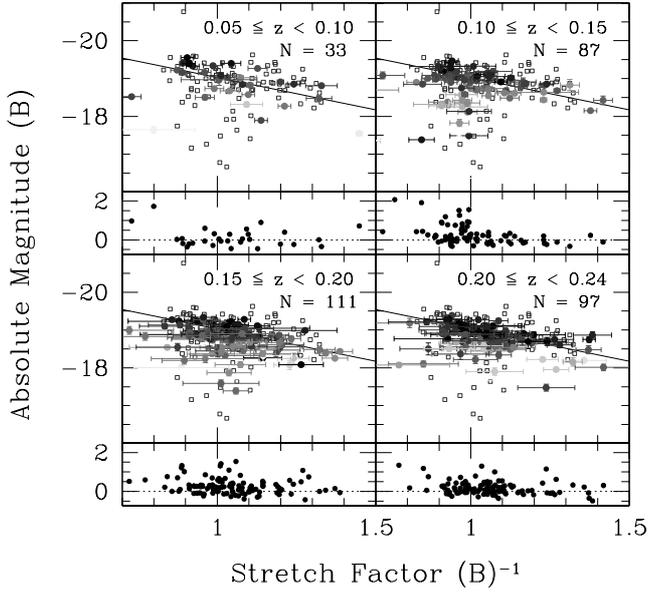}
\end{center}
\caption{Relation between the inverse {\it B-}band stretch factor and {\it B-}band peak magnitude (the upper panels) and $\Delta M_{B}$ (the lower panels) at different redshift bins. In the upper panels, circles are the SDSS sample and open squares are the Nearby sample. The thickness of the circles is related to $(M_{B}-M_{V})_{max}$, the darker circles are the bluer SNe Ia and the lighter circles are the redder SNe Ia (same as Figure \ref{Z_B}). The solid line denotes the relation between the inverse {\it B-}band stretch factor and {\it B-}band magnitude derived from the SDSS sample. In the lower panels, positive $\Delta M_{B}$ indicates that a photometric point is below the regression line.\label{BSF_BMAG_Z}}
\end{figure}

Figure \ref{BSF_BMAG_Z}\footnote{We use inverse {\it B-}band stretch factor in these figures for the comparison with Figure 16 in TAK08.} shows the {\it B-}band stretch-magnitude relation in different redshift bins. 90\% of the SDSS sample become under the stretch-magnitude relation derived from the bluest SDSS sample ($M_{B} = 1.71 \times s^{-1}_{(B)} - 20.73$). 

\begin{table*}
\begin{center}
\begin{minipage}{40em}
\caption{Relations of {\it B-}band stretch factor and {\it B-} and{\it V-}band magnitude of bluest SNe. The relations of Nearby SNe is from TAK08.}
\label{Bsf_mag_coe}
\end{minipage}
\small
\begin{tabular}{cccc} \hline\hline
Sample & Relation & r.m.s. (mag) & Number \\ \hline
Nearby & $M_B = (2.28 \pm 0.42) \times s_{(B)}^{-1} - (21.49 \pm 0.41)$ & 0.17 & 21 \\
 & $M_V = (2.58 \pm 0.38) \times s_{(B)}^{-1} - (21.67 \pm 0.38)$ & 0.20 & 21 \\
SDSS & $M_B = (1.71 \pm 0.31) \times s_{(B)}^{-1} - (20.73 \pm 0.31)$ & 0.16 & 15 \\
 & $M_V = (1.59 \pm 0.26) \times s_{(B)}^{-1} - (20.51 \pm 0.25)$ & 0.16 & 15 \\ \hline
\end{tabular}\\
\vspace*{0.6em}
\end{center}
\end{table*} %

\begin{figure}
\begin{center}
\includegraphics[scale=0.45]{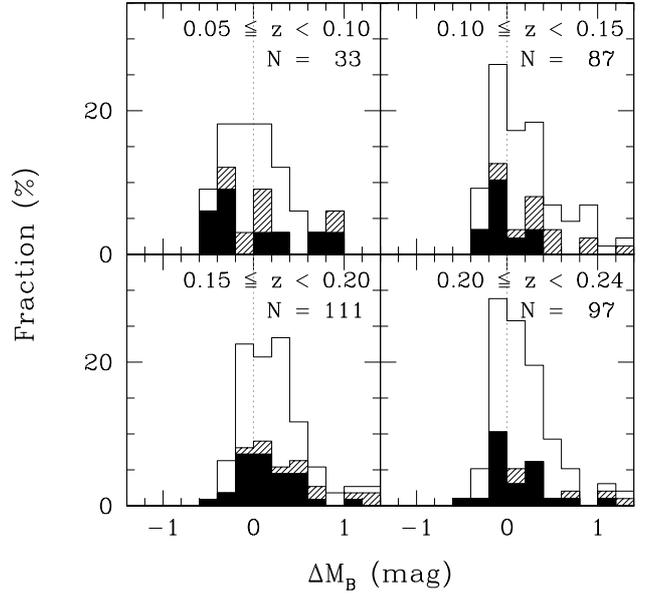}
\end{center}
\caption{Histogram of $\Delta M_{B}$ of the SDSS sample at each redshift bin. The shaded histograms show the Broad SNe Ia, the filled histograms show the Narrow SNe Ia, and the remaining histograms are the Normal SNe Ia. \label{BMAG_RES_HIST}}
\end{figure}

We define the {\it B-}band residual (hereafter $\Delta M_{B}$) as $\Delta M_{B} = M_{B} - (1.71 \times s^{-1}_{(B)} - 20.73)$, which is difference between estimated $M_{B}$ from photometry and intrinsic luminosity estimated from the stretch-magnitude relation of the SDSS sample. Figure \ref{BMAG_RES_HIST} shows the histogram of $\Delta M_{B}$ of the SDSS sample using the stretch-magnitude relation derived from the SDSS sample. Host galaxy dust extinction is one of the possible explanations of the tails toward large $\Delta M_{B}$.

\subsubsection{Stretch-Colour Relation}

\begin{figure}
\begin{center}
\includegraphics[scale=0.45]{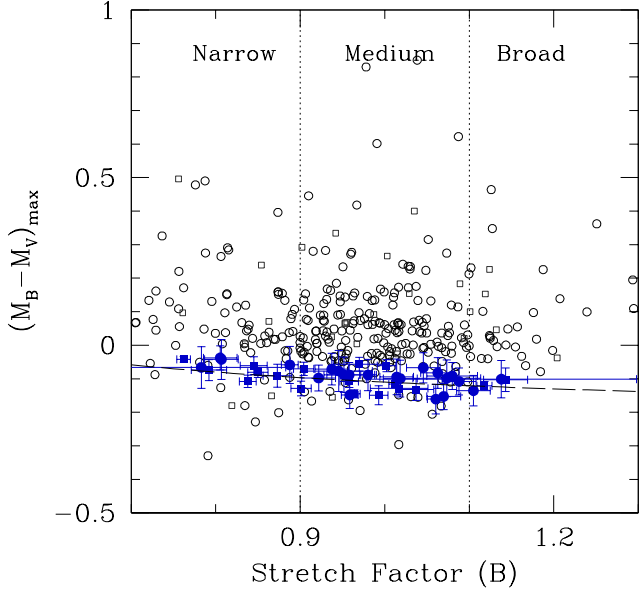}
\end{center}
\caption{A relation between the {\it B-}band stretch factor and $(M_{B}-M_{V})_{max}$ of the Nearby and SDSS sample. The squares are the Nearby SNe Ia and the circles are the SDSS SNe Ia, in which blue filled symbols are the BV-selected SNe Ia with a smaller photometric uncertainty. The diagonal dashed line shows the relation between {\it B-}band stretch factor and $(M_{B}-M_{V})_{max}$ of bluest sample of the SDSS SNe Ia used for the definition of SNCE.  \label{BSF_BV}}
\end{figure}

In order to discuss nature of SNe Ia colour and variety of host galaxy dust extinction at \S6, we use $M_{B} - M_{V}$ colour at the {\it B-}band maximum date because {\it B-} and {\it V-}bands are the bands with the least dispersion among optical passbands (\citealt{jam06}). Figure \ref{BSF_BV} show the colour distribution of the SDSS sample. 93\% of SNe are above the stretch-colour relation derived from the SDSS sample.

We define the supernova colour excess (hereafter SNCE) as $SNCE = (M_{B}-M_{V})_{max} - (0.12 \times s^{-1}_{(B)} - 0.22$), which is difference between estimated $(M_{B}-M_{V})_{max}$ from photometry and intrinsic colour estimated from stretch-magnitude relations of the SDSS sample (Table \ref{Bsf_mag_coe}).

\subsection{Photometric Properties of Host Galaxies}

\begin{table*}
\begin{center}
\begin{minipage}{310pt}
\caption{Relations between $\Delta M_{B}$ and SNCE of each subsample. $R_{V}$ is conversion factor of {\it V-}band translated from $R_{B}$. See also Figure \ref{DBMAG_BV}. \vspace{4pt}}
\label{DBMAG_BV_TABLE}
\end{minipage} \\
\begin{tabular}{cccc} \hline\hline
Sample & Relation & $R_{V}$ & Number \\ \hline
Broad & $(0.18 \pm 0.01) \times \Delta M_{B} + (0.10 \pm 0.01)$ & $4.3^{+0.2}_{-0.3}$ & \phantom{0}39 \\ 
(Blue Host only) & $(0.21 \pm 0.01) \times \Delta M_{B} + (0.09 \pm 0.01)$ & $3.7^{+0.1}_{-0.2}$ & \phantom{0}34 \\ 
Medium & $(0.23 \pm 0.01) \times \Delta M_{B} + (0.10 \pm 0.01)$ & $3.3^{+0.2}_{-0.1}$ & 208 \\ 
(Blue Host only) & $(0.21 \pm 0.02) \times \Delta M_{B} + (0.09 \pm 0.01)$ & $3.7^{+0.3}_{-0.4}$ & 154 \\ 
(Red Host only)& $(0.38 \pm 0.03) \times \Delta M_{B} + (0.14 \pm 0.01)$ & $2.0^{+0.2}_{-0.1}$ & \phantom{0}28 \\ 
Narrow & $(0.19 \pm 0.03) \times \Delta M_{B} + (0.14 \pm 0.01)$ & $4.0^{+0.8}_{-0.5}$ & \phantom{0}81 \\
(Blue Host only) & $(0.34 \pm 0.04) \times \Delta M_{B} + (0.14 \pm 0.01)$ & $2.3^{+0.3}_{-0.3}$ & \phantom{0}49 \\ 
(Red Host only) & $(0.28 \pm 0.13) \times \Delta M_{B} + (0.14 \pm 0.03)$ & $2.8^{+2.3}_{-0.9}$ & \phantom{0}18 \\ \hline 
All & $(0.15 \pm 0.06) \times \Delta M_{B} + (0.12 \pm 0.04)$ & $5.1^{+3.4}_{-1.4}$ & 328 \\ \hline
\end{tabular}
\end{center}
\end{table*}

\subsubsection{Extinction Law of Host Galaxy Dust}

It is important to correct for dust extinction in the host galaxy before making distance estimation using SNe Ia. In particular, we must know the shape of extinction curve (colour-colour relation) and the zeropoint of the extinction curve (magnitude-colour relation).

Since $\Delta M_{B}$ and SNCE are possibly due to host galaxy dust extinction, the $\Delta M_{B}$-SNCE relation gives us a clue to understand the properties of host galaxy dust. Figure \ref{DBMAG_BV} shows the $\Delta M_{B}$-SNCE relation of the SDSS samples which include Medium, Broad and Narrow SNe Ia subsamples. We discuss the possibility that there are different stretch-colour relations for the SNe Ia with different stretch factors in \S6. The relations and conversion factors we derive are shown in Table \ref{DBMAG_BV_TABLE}.

\subsubsection{Relations between SNe Ia and Host Galaxy Colour}

\begin{table*}
\begin{center}
\begin{minipage}{310pt}
\caption{Dispersion of residuals around the relations of $\Delta M_{B}$ and SNCE of each subsample. See also Figure \ref{DBMAG_BV_MATRIX}. \vspace{4pt}}
\label{DBMAG_BV_DIS_TABLE}
\end{minipage} \\
\begin{tabular}{cccccccc} \hline\hline
\multicolumn{2}{c}{Sample} & \multicolumn{2}{c}{Red Host} & \multicolumn{2}{c}{Blue Host} & \multicolumn{2}{c}{All} \\
 & & $\sigma (mag)$ & Number & $\sigma (mag)$ & Number & $\sigma (mag)$ & Number \\ \hline
Broad & A & - & \phantom{0}3 & 0.19 & \phantom{0}34 & 0.19 & \phantom{0}39 \\ 
  & B & - & \phantom{0}2 & 0.04 & \phantom{0}18 & 0.04 & \phantom{0}20 \\ 
Medium & A & 0.08 & 28 & 0.11 & 154 & 0.11 & 208 \\ 
  & B & 0.06 & 22 & 0.07 & 126 & 0.08 & 169 \\ 
Narrow & A & 0.09 & 18 & 0.15 & \phantom{0}49 & 0.12 & \phantom{0}81 \\
  & B & 0.05$^{a}$ & 16 & 0.12 & \phantom{0}43 & 0.11 & \phantom{0}73 \\ \hline
Total & & & 49 & & 237 & & 328 \\ \hline 
\end{tabular}
\vspace*{0.6em} \\
{\footnotesize
\begin{minipage}{310pt}
Sample A is all of SNe of the subgroup.\\
Sample B is SNe with $\Delta M_{B} < 0.5$ mag.\\
a: One outlier is excluded because the SN has larger dispersion than 5 $\sigma$.\\
\end{minipage}}
\end{center}
\end{table*}

We investigated relationships between the photometric properties of SNe Ia and their host galaxy types. Using the galaxies brighter than ${\it g} = 21$ in the SDSS imaging data, \cite{str01} found that 90\% of spectroscopically classified late-type galaxies have colour bluer than ${\it u- r} = 2.5$ and found no examples of spectroscopically classified early-type galaxies with colour bluer than ${\it u - r} = 2.05$. Based on the ${\it u - r}$ colour of the host galaxy, we classified the SDSS SNe Ia into three subgroups as follows. (1) ``Blue Host" : ${\it u - r} < 2.0$ corresponding to late types (Sb, Sc, and Irr), (2) ``Red Host" : ${\it u - r} > 2.5$ corresponding to early types (E/S0 and Sa), and (3) intermediate colour : $2.0 \le {\it u - r} \le 2.5$. We show the relations between SNe Ia and host galaxy type in Table \ref{DBMAG_BV_TABLE}, Figure \ref{DBMAG_BV} and \ref{DBMAG_BV_MATRIX}.

\cite{bla07} and \cite{ski09} show a different colour cut, $g - r = 0.8 - 0.03 \times (M_r + 20)$ for separating the ``red sequence" and ``blue cloud" of galaxies. However, this colour cut is similar to the colour cut proposed by \cite{str01} and the latter is a more strict colour cut for selecting early type galaxies. We use the colour cut proposed by \cite{str01} in this study. The specific star formation rate (sSFR) of 125 ``Blue Host" galaxies and 16 ``Red Host" galaxies are published by \cite{wol16}. The average of sSFR of the ``Blue Host" galaxies is -9.78 $M_{\odot} / yr$ and that of the ``Red Host" galaxies is -11.34 $M_{\odot} / yr$. This result supports the idea that the ``Blue Host" galaxies correspond to late types and the ``Red Host" galaxies correspond to early types.

Figure \ref{DBMAG_BV} shows that the SNe Ia which appeared in the ``Red Hosts" have  smaller dispersion in SNCE around the best fit relation than that of the SNe Ia which appeared in the ``Blue Hosts" (see also Table \ref{DBMAG_BV_DIS_TABLE}). In addition, the Medium and Narrow SNe Ia which appeared in the ``Red Hosts" have smaller $\Delta M_{B}$ than that of the SNe Ia which appeared in the ``Blue Hosts". Table \ref{DBMAG_BV_DIS_TABLE} also shows that the dispersion in SNCE around the best fit relation is smaller in each SN Ia subsample with $\Delta M_{B} < 0.5$ mag (\ref{DBMAG_BV_DIS_TABLE}B).

\section{DISCUSSION}
\label{discussion}

Since the SDSS SN Ia sample is more homogeneous than any other large sample of SNe Ia made from surveys at lower redshift, the photometric properties derived would give us a clue to understand the nature of spectroscopically normal SN Ia. Hence, it is worth looking into photometric properties of SDSS SNe Ia in detail.

In particular, the $\Delta M_{B}$-SNCE relation is quite interesting (Figure \ref{DBMAG_BV}). The dispersion around the $\Delta M_{B}$-SNCE relation (Table \ref{DBMAG_BV_DIS_TABLE}) might be consistent with the uncertainties (see Table \ref{ERROR_TABLE}) under the two assumptions that (1) peak luminosities and colours are the same for all SNe Ia after the stretch correction, and (2) extinction laws of dust in host galaxies are the same. However, the dispersion of the Medium sample (0.11 mag) appears to be larger than the typical  uncertainty ($\sigma$ in colour is 0.08 mag at $z = 0.2$). This implies that the assumptions may be too simplified. We now discuss the possible diversity of photometric properties of SNe Ia and properties of dust extinction in their host galaxies.

\subsection{Diversity of Photometric Properties of SNe Ia}

\begin{figure}
\begin{center}
\includegraphics[scale=0.45]{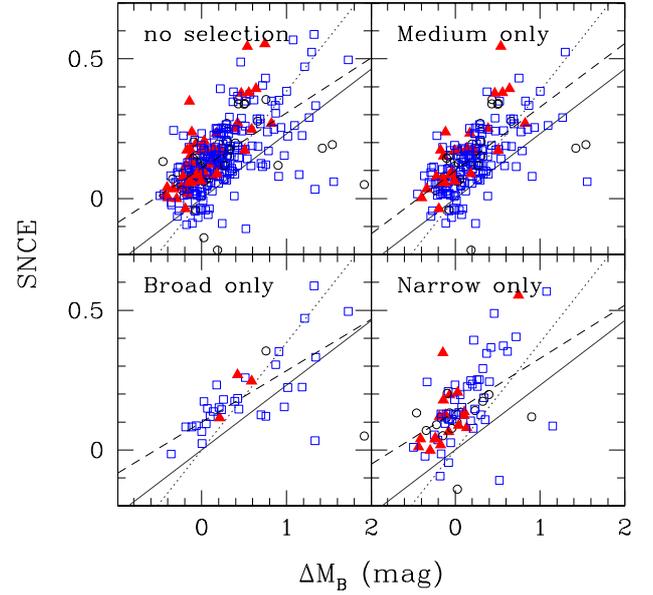}
\end{center}
\caption{Relations between $\Delta M_{B}$ from the stretch-magnitude relation and SNCE from the stretch-colour relation of the SDSS SNe Ia. Top left panel shows all of the SDSS sample and the other panels show subsamples (Medium, Broad, Narrow, respectively). The blue open squares show the SNe Ia which appeared in the ``Blue Host" galaxies, the red filled triangles show the SNe Ia which appeared in the ``Red Host" galaxies, and the black open circles show the SNe Ia which appeared in the intermediate colour galaxies. Average estimation uncertainties are 0.06 mag for $\Delta M_{B}$ and 0.08 mag for SNCE. The solid lines denote the conversion factor of standard Galactic extinction ($R_{V} = 3.3$), the dotted lines denote smaller conversion factor ($R_{V} = 2.0$), and the dashed lines denote the best linear fit. See also Table \ref{DBMAG_BV_TABLE}. \label{DBMAG_BV}}
\end{figure}

\begin{figure}
\begin{center}
\includegraphics[scale=0.45]{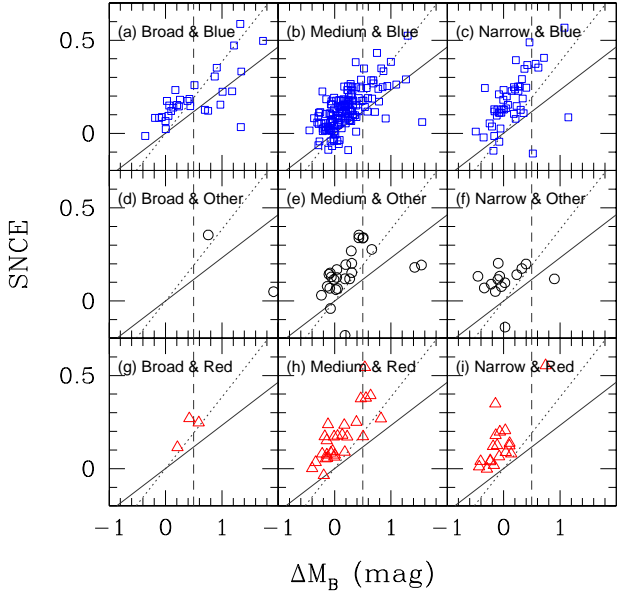}
\end{center}
\caption{Subsets of the relation between $\Delta M_{B}$ and SNCE of the SDSS SNe Ia. The blue open squares show the SNe Ia which appeared in the ``Blue Host" galaxies (a, b, c), the black open circles show  the SNe Ia which appeared in the intermediate colour galaxies (d, e, f) and the red open triangles show the SNe Ia which appeared in the ``Red Host" galaxies (g, h, i). Left side of the dashed line is the ``B" sample (see Table \ref{DBMAG_BV_DIS_TABLE}). They are also divided into three subgroups, Broad, Medium and Narrow. The solid line is the conversion factor of standard Galactic extinction ($R_{V} = 3.3$) and the dotted line denotes smaller conversion factor ($R_{V} = 2.0$). \label{DBMAG_BV_MATRIX}}
\end{figure}

Table \ref{DBMAG_BV_DIS_TABLE} shows that the SNe Ia which appeared in the ``Red Hosts" tend to have narrower light curve width (3 / 39 SNe of the Broad sample, 28 / 208 SNe of the Medium sample and 18 / 81 SNe of the Narrow sample are ``Red Hosts" SNe; see also the ratio of red filled triangles to others in Figure \ref{DBMAG_BV}). Since many of ``Red Hosts" are early type galaxies, the tendency that the ``Red Hosts" SNe Ia have a narrower light curve width is consistent with previous studies (c.f. \citealt{ham96b,ham00,how01,van05,sul06}). The SNe Ia which appeared in the ``Blue Hosts", however, have a variety of light curve widths  (34 / 39 SNe of the Broad sample, 154 / 208 SNe of the Medium sample and 49 / 81 SNe of the Narrow sample are ``Blue Hosts" SNe; see also the ratio of blue open squares to others in Figure \ref{DBMAG_BV}).

These observational results can be explained by postulating that the progenitors of SNe Ia which appeared in the ``Red Hosts" are members of an old stellar population, but those of SNe Ia which appeared in the ``Blue Hosts" may belong to both old and young stellar populations. However, there are 6 Medium SNe Ia which have large $\Delta M_{B}$ in spite of they appeared in the ``Red Hosts" (see top right panel of Figure \ref{DBMAG_BV}). They can be explained if those ``Red Hosts" are dusty red galaxies with large extinction. In fact, \cite{str01} reported about 10\% (20/210) of late-type galaxies have a colour of $u - r > 2.5$. Table \ref{LARGE_DBMAG_TABLE} shows some of those ``Red Hosts" may be dusty red late-type galaxies.

\begin{table*}
\begin{center}
\begin{minipage}{350pt}
\caption{Host galaxies of SNe Ia with $\Delta M_{B} \geq 0.5$ which are excluded from sample B in Table \ref{DBMAG_BV_DIS_TABLE}. Morphology is determined by eye inspections and spectral type is determined by SDSS-III's Baryon Oscillation Spectroscopic Survey (BOSS, Dawson et al. 2013). \vspace{4pt}}
\label{LARGE_DBMAG_TABLE}
\end{minipage} \\
\begin{tabular}{cccccc} \hline\hline
$\Delta M_{B}$ & SDSS-SNID & SN name & Host name & Morphology & Spectral Type \\ \hline
0.51 & 13224 & - & SDSS J030958.79-001444.8 & ? & N/A \\ 
0.54 & 20376 & - & SDSS J211734.92-003126.3 & ? & N/A \\ 
0.56 & 12778 & 2007re & SDSS J210958.95+002430.9 & Spiral & Starforming \\ 
0.64 & 1415 & - & SDSS J002425.55+003556.4 & Spiral & Broadline \\ 
0.82 & 3488 & - & SDSS J205413.25-010037.2 & ? & N/A \\ 
2.54 & 19003 & 2007mp & 2MASX J21163594-0046133 & Spiral & AGN Broadline \\ \hline
\end{tabular}
\end{center}
\end{table*}

The interpretation that there are two types of SNe Ia with different photometric properties, originating in young and old stellar populations, is consistent with the idea that there are two varieties of SNe Ia, named ``tardy'' and ``prompt'' (\citealt{man05}). The SNe Ia which appear in the ``Red Hosts" are all ``tardy'' SNe Ia and the SNe Ia which appear in the ``Blue Hosts" are both ``tardy'' and ``prompt'' SNe Ia. Based on this idea, we can estimate the intrinsic dispersions of the ``tardy'' and ``prompt'' SNe Ia on the $\Delta M_{B} - SNCE$ plane.  In Table \ref{DBMAG_BV_DIS_TABLE}, the B samples of ``Red Host" - ``Narrow" and  ``Red Host" - ``Medium" are expected to be pure samples of ``tardy" SNe Ia and the B sample of ``Blue Host" - ``Broad" is ``prompt" SN Ia. Those dispersions are 0.04-0.06 mag, and the values are smaller than that of the B samples of ``Blue Host" - ``Medium" (0.07 mag) and ``Blue Host" - ``Narrow" (0.12 mag) SNe Ia which are expected to be a mixture of ``tardy'' and ``prompt'' SNe Ia. This result supports the original idea that there are two distinct population (``tardy'' and ``prompt'') proposed by \cite{man05}.

\begin{figure}
\begin{center}
\includegraphics[scale=0.45]{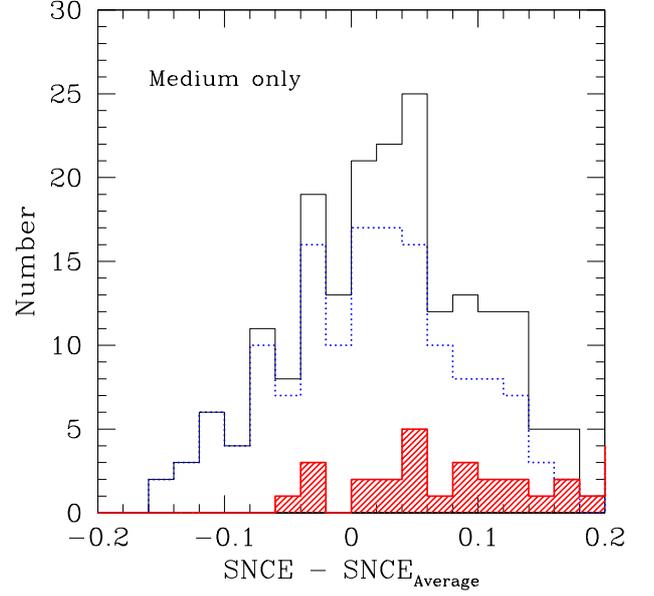}
\end{center}
\caption{Distributions of colour residuals from the average SNCE (dashed line in Figure \ref{DBMAG_BV}) of the Medium SNe Ia. The histogram with a dotted blue line shows the ``Blue Host" SNe Ia, the shaded histogram with a red line shows the ``Red Host" SNe Ia and the histogram with a solid black line shows all of the Medium SNe Ia. Averages of colour residuals are 0.01 mag for the ``Blue Host" SNe Ia  and 0.07 mag for  the ``Red Host" SNe Ia.  \label{EBV_HOST_HIST}}
\end{figure}

Figure \ref{EBV_HOST_HIST} shows the distributions of SNCE residuals from the regression line of $\Delta M_{B} - SNCE$ plane of the Medium SNe Ia. The result of Kolmogorov-Smirnov test shows that the distributions of the Medium SNe Ia which appeared in the ``Blue Hosts" and ``Red Hosts" are different at a significance level 99.9\%. Averages of colour residuals are 0.01 mag for the ``Blue Host" SNe Ia  and 0.07 mag for the ``Red Host" SNe Ia. The colour offset,  ``tardy'' SNe are redder and ``prompt'' SNe are bluer, can be explained by the idea of two types of SNe Ia with different intrinsic colour. These results support the possibility that there are two different populations of SNe Ia.

\begin{figure}
\begin{center}
\includegraphics[scale=0.45]{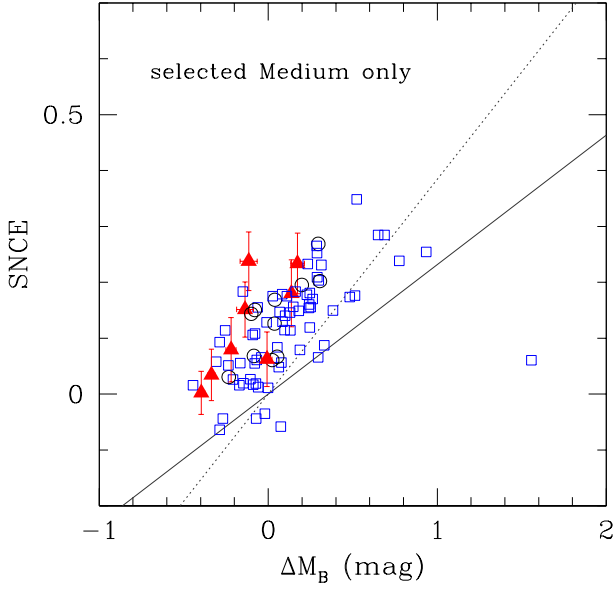}
\end{center}
\caption{A relation between $\Delta M_{B}$ from the stretch-magnitude relation and SNCE from the stretch-colour relation of the selected Medium SDSS SNe Ia with small uncertainties ($0.9 < s_{(B)} \le 1.1$, $\sigma_{M_{B}}, \sigma_{M_{V}} < 0.05$ mag).  The symbols and lines are the same as Figure \ref{DBMAG_BV} (error bars are added to the ``Red Host" SNe Ia). \label{DBMAG_BV_SELECTED_GAL}}
\end{figure}

\begin{figure}
\begin{center}
\includegraphics[scale=0.45]{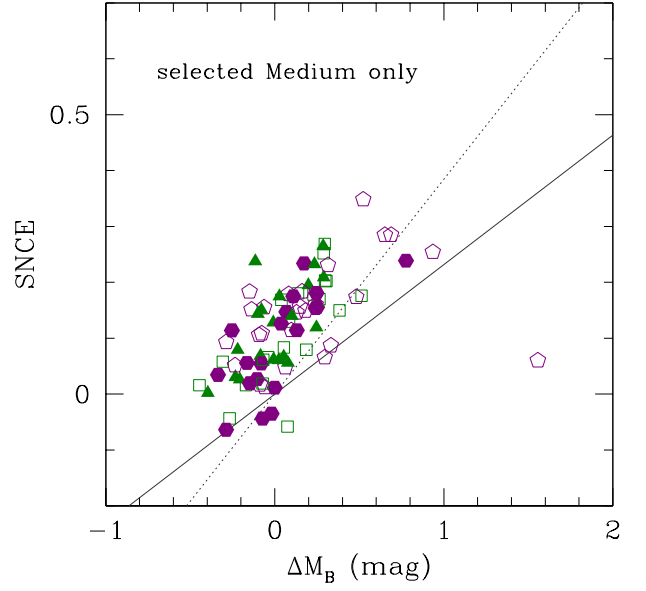}
\end{center}
\caption{A relation between $\Delta M_{B}$ and SNCE of the selected Medium SDSS SNe Ia which divided into four groups due to a stretch factor. Green filled triangles are the SNe Ia with $0.90 < s_{(B)} \le 0.95$, green squares are the SNe Ia with $0.95 < s_{(B)} \le 1.00$, purple pentagons are the SNe Ia with $1.00 < s_{(B)} \le 1.05$, and purple filled sexanglulars are the SNe Ia with $1.05 < s_{(B)} \le 1.10$. The lines are the same as Figure \ref{DBMAG_BV}. \label{DBMAG_BV_SELECTED_SF}}
\end{figure}

There are several possibilities which could mislead this analysis. The first possibility is large estimated uncertainties. However, the dispersion of colour residuals is significantly larger than the estimated uncertainties even if we select the Medium SNe Ia with smaller uncertainties ($\sigma_{M_{B}}, \sigma_{M_{V}} < 0.05$ mag, see Figure \ref{DBMAG_BV_SELECTED_GAL}). The second possibility is that the two types of SNe Ia correspond to the Medium SNe Ia with larger $s_{(B)}$ and smaller $s_{(B)}$, because the $s_{(B)}$ range of the Medium SNe Ia is too wide to regard it as a homogeneous sample. However, the tendency is the same for the Medium SN Ia subsamples divided into four groups based on the {\it B-}band stretch factor ($0.90 \le s_{(B)} < 0.95, 0.95 \le s_{(B)} < 1.00, 1.00 \le s_{(B)} < 1.05, 1.05 \le s_{(B)} < 1.10$, see Figure \ref{DBMAG_BV_SELECTED_SF}). The result of Kolmogorov-Smirnov test shows that the distributions of the Medium SNe Ia with smaller $s_{(B)} (0.90 \le s_{(B)} < 0.95)$ and that of the Medium SNe Ia with larger $s_{(B)} (1.05 \le s_{(B)} < 1.10)$ are not different (p-value = 0.28). The third possibility is that the inverse stretch-magnitude relation ($M_{B} = 1.71 \times s^{-1}_{(B)} - 20.73$), which is used for deriving $\Delta M_{B}$ creates apparent two subgroups. However, we confirmed that there is no significant difference ($\sim 1\%$) between the Medium SN Ia subsamples based on the inverse and normal stretch-magnitude relation.

From these discussions, we conclude there may be two subgroups of Medium SNe Ia which have intrinsically different colours.

\begin{table}
\begin{center}
\begin{minipage}{180pt}
\caption{Difference of properties of host galaxies between the SNLS sample used in Sullivan et al. (2010) and the SDSS sample. The average values of the SNLS sample are calculated from Table 1 except metallicity which is read from Figure 11 of the paper.  \vspace{4pt}}
\label{COMP_W_SUL}
\end{minipage} 
\begin{tabular}{lccc} \hline\hline
 & SNLS & this work \\ \hline
redshift & 0.64 & 0.16 \\
stellar mass [log M$_\odot$] & 9.8 & 10.5 \\
metallicity [12+log(O/H)] & 8.5 & 8.9 \\
sSFR [log M$_\odot$/yr] & -9.5 & -10.0 \\ \hline
\end{tabular} \\
\end{center}
\end{table} %

\cite{sul10} examined the colour-magnitude relation of the SNLS SNe sample. Although Figure 9 of \cite{sul10} shows basically the same properties as those in Figure \ref{DBMAG_BV}, we cannot identify two distinct subgroups as in the Medium SDSS sample (right top panel of Figure \ref{DBMAG_BV}). However, their results do not conflict with our findings since we cannot identify two subgroups clearly in the SDSS sample which include the Narrow, Medium, and Broad SNe (Figure \ref{DBMAG_BV}). There is another possible explanation that difference of redshift range of the samples may affect the results. Redshift of the SNLS SNe sample are higher than that of our sample, and properties of host galaxies such as stellar mass, metallicity and sSFR are different between the samples (see Table \ref{COMP_W_SUL}). Two distinct subgroups in the Medium SDSS sample may be a result of a variety of stellar population.

\cite{lam10b} discussed the relation between properties of SN Ia and their host galaxies in SDSS sample which includes 110 common SNe with this study. They use the SALT2 (\citealt{guy07}) and the MLCS2k2 (\citealt{jha07}) for the parameterization of multi-band light curves, and found that introduction of the third parameter, a host galaxy type, reduces the luminosity dispersion after the corrections. \cite{bra10} also found strong evidence of two progenitor channels related with the age of SN Ia progenitors from the delay time distribution of SDSS sample. They also use the SALT2 (\citealt{guy07}) for the parameterization of multi-band light curves, and found ``tardy'' components are luminous SNe Ia with high stretch factors and ``prompt'' components are subluminous SNe Ia with low stretch factors. Even though we use a different light curve analysis method as we described \S3, these results are consistent with our conclusion that there are two subgroups of SNe Ia depending on the host galaxy colour. 

\subsection{Extinction Law of Host Galaxy Dust}

Dust is another possible factor for creating large dispersions in the colour distributions of SNe Ia which appear in the ``Blue Hosts" and ``Red Hosts". The colour-colour relations derived from the SDSS sample indicate that the shape of extinction curves are similar to that of the Galactic dust. On the other hand, the relations between $\Delta M_{B}$ and SNCE, which is related to the conversion factor $R$, have different implications. 

From Figure \ref{DBMAG_BV}, it is not clear whether or not the average $R_{V}$ of host galaxy dust is close to the standard Galactic value ($R_{V} = 3.3$) . As shown in $\S6.1$, the SDSS sample may include two types of SNe Ia which have intrinsically different colours. In order to reduce the uncertainty of intrinsic diversity of SNe Ia colour, we derive $R_{V}$ from the Medium, Broad and Narrow samples. As a result, we find that a value of $R_{V}$ derived from the Medium sample is $3.3^{+0.2}_{-0.1}$, and it is consistent with the standard Galactic value. On the other hand, values of $R_{V}$ derived from the Medium-``Red Hosts" sample ($2.0^{+0.2}_{-0.1}$) and Narrow-``Blue Hosts" sample ($2.3^{+0.3}_{-0.3}$) are significantly smaller than the standard Galactic value. These results are consistent with previous studies (e.g. $R_{B} \sim 3.5$, \citealt{phi99,kno03,alt04}; $R_{B} \sim 3.3$, \citealt{wan06}; $R_{V} = 1.75 \pm 0.27$, \citealt{nob08}; $R_{V} = 2.2$, \citealt{kes10}), which find values of $R$ are smaller than the standard Galactic value. 

These mesurements may indicate that the value of $R_{V}$ has a large intrinsic dispersion. Also we should notice the report that there are large variations of interstellar extinction curve even in our Galaxy (\citealt{nat15}). An example is given by \citet{folatelli10}, who found $R_V \approx 1.7$ for their sample of SNe Ia but this value changed to $R_V \approx 3.2$ if the highly reddened SNe are excluded. 

\subsection{Other factors which may have affected the observed SNe Ia colours}

In the line of sight toward any SN Ia, there may be circumstellar dust (CSD) around the SN Ia in addition to dust in the interstellar medium (ISM). The existence of circumstellar material (CSM) is expected based on theoretical arguments (e.g. \citealt{nom82,hac99}), and several recent observations have indicated the existence of circumstellar gas around spectroscopically normal SNe Ia (SN2000cx, \citealt{maz05,pat07a}; SN2002ic, \citealt{ham03}; SN2003du, \citealt{ger04}; SN2005cg, \citealt{qui06}; SN2006X, \citealt{pat07b}; SN2007af, \citealt{sim07}; SN2012ca and SN2013dn, \citealt{fox15}). \cite{wan05} reported a possibility of CSD associated with circumstellar gas has smaller value of $R_{V}$ than that of the Galactic ISM. Smaller $R$ values normally indicate a smaller size of dust particles (c.f. \citealt{dra03}). \cite{wan05} also pointed out that a low $R_{V}$ of host galaxy dust estimated from SNe Ia may be a result of the reflection effect from CSD. The distribution that the SNe appeared in the ``Red Hosts" (red triangles in Figure \ref{DBMAG_BV}) are plotted along a low $R_{V}$ line can be explained by these ideas.

New observations provide increasing number of evidence suggesting the presence of CSM/CSD in (at least some of) SNe Ia. \cite{blo09} detected variable Na I D lines in low-resolution spectra of SN1999cl and SN2006X, suggesting that the variability could be associated with interstellar absorption. \citet{sternberg11} showed that the sodium absorption features in 35 SNe Ia they observed indicate the presence of CSM around the progenitor system, which may have originated from gas outflows from the single-degenerate progenitors. Later, \citet{dilday12} reported a complex, multi-shell circumstellar material structure in the close environment of SN Ia PTF 11kx. An RS Oph-like symbiotic nova progenitor was proposed for this particular SN. Further, \citet{forster12,forster13} put forward new evidences for the presence and asymmetric distribution of CSM in SNe Ia from the observations of nearby SNe Ia. Also recently, \citet{johansson13} obtained strict upper limits for the amount of dust around three nearby SNe Ia down to $M_{dust}\lesssim7\times10^{-3}$ M$_\odot$ with Herschel Space Observatory far-infrared observations, but could not completely rule out CSD as one contributor to the reddening suffered by SNe Ia.

Another possible factor affecting the observed SN Ia colour is the viewing angle of the asymmetric SN explosion. An off-center explosion of the white dwarf progenitor star in a SN Ia \citep{maeda10} may affect the  luminosity and colour of the SN \citep{maeda11,cartier11}. SNe viewed from the side closer to the ignition center will appear bluer, and those viewed from the other side will appear redder. \citet{maeda11} shows a difference of the peak luminosity changes $\sim 30\%$ due to the viewing angle. The colour of the SN also seems to be correlated with the ejecta velocity \citep{foley11,foley12}, with high-velocity SNe tend to be redder than the low-velocity ones and the offset is $\sim0.04$ mag. As we discussed in \S6.1, if the SNe appeared in ``Red Host'' except dust rich late-type galaxies are all ``tardy'' SNe Ia, the viewing angle is one of possible explanations of their distribution in Figure \ref{DBMAG_BV}. On the other hand, we cannot conclude that dependence on the ejecta velocity are seen  in our sample.

\begin{figure}
\begin{center}
\includegraphics[scale=0.45]{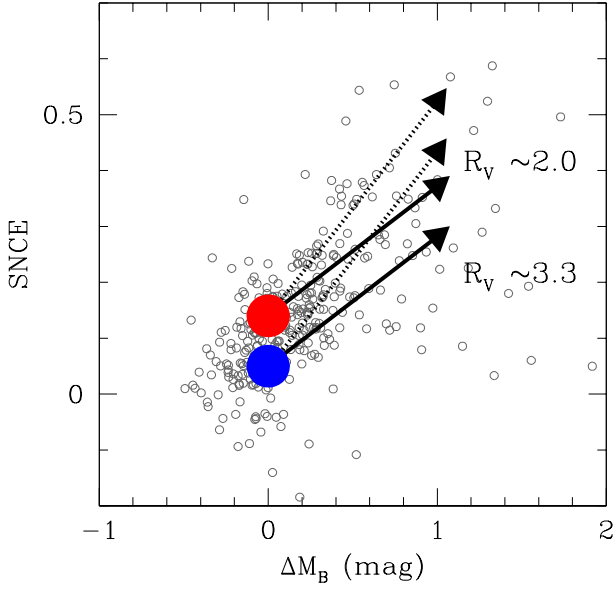}
\end{center}
\caption{A schematic picture of our conclusion based on the discussion in \S6. There are two types of SNe Ia with different intrinsic colours (the circles), and they are affected by dust with different extinction laws (the arrows). The solid line denotes the conversion factor of standard Galactic extinction ($R_{V} = 3.3$) and the dotted line denotes smaller conversion factor ($R_{V} = 2.0$). Grey small circles are the SDSS sample same as top left panel of Figure \ref{DBMAG_BV}, which probably includes a mixture of these factors. \label{DBMAG_BV_DEF}}
\end{figure}

While with the current photometric dataset we are unable to address the aforementioned points affecting the colours of our SNe, disregarding those possible factors we present the empirical result of our study as follows. A schematic picture of our conclusion based on the discussion above is shown in Figure \ref{DBMAG_BV_DEF}. There may be two types of SNe Ia with different intrinsic colours, and their colours are affected by dust with two different extinction properties.

This idea does not conflict with \cite{sco14b} which pointed out the Galactic reddening law ($R_v = 3.1$) can explain the trend between Hubble residuals and colours. As we show in Table \ref{DBMAG_BV_TABLE}, the reddening law ($R_v = 3.3$) derived from the Medium sample consistent with the Galactic reddening law. The smaller value of $R_v$ can be derived if we regard the SN sample as two components as we discussed. 

\subsection{Implications for cosmology}

If there are two subgroups of Medium SNe Ia which have intrinsically different colours, and they are extincted by host galaxy dust with different properties, what are implications for cosmological studies? 

The offset in intrinsic colour between two subgroups of Medium SNe Ia is 0.06 mag as we showed in \S6.1. This colour offset becomes $\Delta M_B = 0.26$ mag under the standard galactic extinction law ($R_B = 4.3$). This is significantly large value compared with the standard value of peak brightness dispersion $\sigma_{M_B} \sim 0.15$. However, since the Medium SNe appeared in ``Red Host'' are only 13\% of all the Medium SNe (28 / 208), the large offset is smoothed when we use a SN sample with no selection bias. On the other hand, when we use a SNe sample selected by host galaxy type, it should be carefully analyzed since the offset may become a factor of systematic uncertainty. If there are statistically enough number of the Medium SNe appeared in ``Red Host'', dispersion of peak brightness is expected to be smaller than that of the all Medium SNe.

In order to reduce uncertainty due to variety of $R_{V}$, we should use only the bluest SNe Ia for cosmological studies unless we know a proper $R_{V}$ of each supernova. However, it is needed to carefully select the bluest SNe Ia subsample when we use them. If we assume that a SN with $-0.2 \le M_B-M_V \le -0.1$ is the bluest one, there are 13\% of all the Medium SNe (28 / 208). This colour selected sample reduces systematic uncertainties in luminosity estimation by factor 2, but the small number of the sample results in increasing statistical uncertainties. As we show in Table \ref{DBMAG_BV_DIS_TABLE}, the effects are cancelled in this study (Dispersion of Medium-B subsamples is similar in spite of difference in sample size). 

When we have larger SN Ia sample in future, using subsamples of carefully selected SNe Ia may improve precision of the cosmological measurements with SNe Ia. Cosmological studies are not the main topic in this paper, but current study may become a clue for the optimized sample selection.

\section{SUMMARY}
\label{summary}

In this paper, we present photometric properties of the intermediate redshift SNe Ia found by SDSS-II SN Survey. The {\it u-, g-, r-, i-} and {\it z}-band light curves of the SDSS SNe Ia are parameterized into rest-frame {\it U-, B-, V-, R-} and {\it I-}band stretch factors, peak luminosities, and {\it B-}band maximum luminosity dates using the Multi-band Stretch Method, which can simply parameterize light curve shapes and peak brightness without any additional assumptions such as dust extinction models unlike widely used methods SALT2 and MLCS.
 
Considering the observational uncertainties and selection biases, we select 328 SNe Ia in the range $0.05 \le z < 0.24$ for a study of their photometric properties. Applying the stretch-magnitude and stretch-colour relations derived from the SDSS sample, we find that most of the SNe Ia which appeared in the ``Red Hosts" do not have a broad light curve, on the other hand, the SNe Ia which appeared in the ``Blue Hosts" have a variety of light curve widths. The Kolmogorov-Smirnov test shows that SNe Ia of these two subsamples selected by the host galaxy colour have significantly different colour (significance level of 99.9\%). 

We infer that ``tardy" SNe Ia appeared in both ``Red Hosts" and ``Blue Hosts" but ``prompt" SNe Ia appeared in only ``Blue Hosts". Based on the inference, the Medium and Narrow SNe Ia appeared in ``Red Hosts" are pure ``tardy" samples, and the Broad SNe Ia appeared in ``Blue Hosts" are pure ``prompt" samples. Both of these pure samples have a smaller dispersion ($\le 0.06$ mag) in the supernova colour excess (SNCE) around the best fit relation than that of the Medium and Narrow SNe Ia appeared in ``Blue Hosts" which are expected to be a mixture of ``tardy" and ``prompt" SNe Ia. 

We also investigate the extinction law of host galaxy dust. As a result, we find that a value of $R_{V}$ derived from the Medium sample is $3.3^{+0.2}_{-0.1}$, and it is consistent with the standard Galactic value. On the other hand, values of $R_{V}$ derived from the Medium-``Red Hosts" sample ($2.0^{+0.2}_{-0.1}$) and Narrow-``Blue Hosts" sample ($2.3^{+0.3}_{-0.3}$) are significantly smaller than the standard Galactic value.

These results indicate that there may be two types of SNe Ia with different intrinsic colours, and they are extincted by host galaxy dust with two different properties. This idea does not claim a review of cosmological studies with unbiased SNe Ia sample. However, there is a possibility that the systematic uncertainty could be reduced by selecting SNe Ia based on the idea.

\section*{Acknowledgments}

Funding for the creation and distribution of the SDSS and SDSS-II has been provided by the Alfred P. Sloan Foundation, the Participating Institutions, the National Science Foundation, the U.S. Department of Energy, the National Aeronautics and Space Administration, the Japanese Monbukagakusho, the Max Planck Society, and the Higher Education Funding Council for England. The SDSS Web site is http://www.sdss.org/. This work was also supported in part by a JSPS core-to-core program ``International Research Network for Dark Energy'' and by a JSPS research grants. We also acknowledges support by CONICYT through FONDECYT grant 3140563, and by Project IC120009 ``Millennium Institute of Astrophysics (MAS)" of the Iniciativa Cient\'ifica Milenio del Ministerio de Econom\'ia, Fomento y Turismo de Chile.

The SDSS is managed by the Astrophysical Research Consortium for the Participating Institutions. The Participating Institutions are the American Museum of Natural History, Astrophysical Institute Potsdam, University of Basel, Cambridge University, Case Western Reserve University, University of Chicago, Drexel University, Fermilab, the Institute for Advanced Study, the Japan Participation Group, Johns Hopkins University, the Joint Institute for Nuclear Astrophysics, the Kavli Institute for Particle Astrophysics and Cosmology, the Korean Scientist Group, the Chinese Academy of Sciences (LAMOST), Los Alamos National Laboratory, the Max-Planck-Institute for Astronomy (MPA), the Max-Planck-Institute for Astrophysics (MPiA), New Mexico State University, Ohio State University, University of Pittsburgh, University of Portsmouth, Princeton University, the United States Naval Observatory, and the University of Washington.

This work is based in part on observations made at the following telescopes. The Hobby-Eberly Telescope (HET) is a joint project of the University of Texas at Austin, the Pennsylvania State University, Stanford University, Ludwig-Maximillians-Universit\"at M\"unchen, and Georg-August-Universit\"at G\"ottingen. The HET is named in honor of its principal benefactors, William P. Hobby and Robert E. Eberly. The Marcario Low-Resolution Spectrograph is named for Mike Marcario of High Lonesome Optics, who fabricated several optical elements for the instrument but died before its completion; it is a joint project of the Hobby-Eberly Telescope partnership and the Instituto de Astronom\'ia de la Universidad Nacional Aut\'onoma de M\'exico. The Apache Point Observatory 3.5 m telescope is owned and operated by the Astrophysical Research Consortium. We thank the observatory director, Suzanne Hawley, and site manager, Bruce Gillespie, for their support of this project. The Subaru Telescope is operated by the National Astronomical Observatory of Japan. The William Herschel Telescope is operated by the Isaac Newton Group on the island of La Palma in the Spanish Observatorio del Roque de los Muchachos of the Instituto de Astrofisica de Canarias. The W. M. Keck Observatory is operated as a scientific partnership among the California Institute of Technology, the University of California, and the National Aeronautics and Space Administration. The Observatory was made possible by the generous financial support of the W. M. Keck Foundation.

We thank the SDSS collaborators for the discussions and meaningful suggestions for this work. We also thank Michael Richmond for carefully reading the manuscript.

\label{biblio}

\clearpage

\clearpage

\label{lastpage}

\end{document}